\documentclass[a4paper,11pt]{article}
\pdfoutput=1
\usepackage{dcolumn}% Align table columns on decimal point
\usepackage{bm}% bold math
\usepackage[T1]{fontenc}%for bold smallcaps
\usepackage{graphicx}% Include figure files
\usepackage{amssymb,amsmath}
\usepackage{multirow}
\usepackage{cite,color,url}
\usepackage[dvipsnames]{xcolor}
\def\linkcolor{cyan!70!black}

\usepackage[
colorlinks=true
,urlcolor=\linkcolor
,anchorcolor=\linkcolor
,citecolor=\linkcolor
,filecolor=\linkcolor
,linkcolor=\linkcolor
,menucolor=\linkcolor
,linktocpage=true
,pdfproducer=medialab
,pdfa=true
]{hyperref}

\usepackage{slashed}
\usepackage{epsfig,psfrag,rotating,soul}
\usepackage{rotfloat}
\usepackage[font={small}]{caption}
\usepackage{colortbl}
\usepackage{soul}
\evensidemargin \oddsidemargin
\newcommand{\be}{\begin{equation}}
\newcommand{\ee}{\end{equation}}

\newcommand{\MET}{E_T^\text{miss}}

\newcommand{\beq}{\begin{equation}} 
\newcommand{\eeq}{\end{equation}} 
\newcommand{\ba}{\begin{array}}  
\newcommand{\ea}{\end{array}} 
\newcommand{\bea}{\begin{eqnarray}}  
\newcommand{\eea}{\end{eqnarray} }  
\newcommand{\bal}{\begin{align}}
\newcommand{\eal}{\end{align}}   
\newcommand{\bi}{\begin{itemize}}  
\newcommand{\ei}{\end{itemize}}  
\newcommand{\ben}{\begin{enumerate}}  
\newcommand{\een}{\end{enumerate}}  
\newcommand{\bc}{\begin{center}}
\newcommand{\ec}{\end{center}} %https://www.overleaf.com/project/6385ef020ce157e09c05bd13
\newcommand{\bt}{\begin{table}}
\newcommand{\et}{\end{table}}  
\newcommand{\btb}{\begin{tabular}}
\newcommand{\etb}{\end{tabular}}

\allowdisplaybreaks

\let\OLDthebibliography\thebibliography
\renewcommand\thebibliography[1]{
  \OLDthebibliography{#1}
  \setlength{\parskip}{0pt}
  \setlength{\itemsep}{0pt plus 0.3ex}
}

\usepackage{fontawesome5}
\makeatletter
\newcommand{\github}[1]{%
   \href{#1}{\faGithubSquare}%
}
\makeatother

\begin{document}

\begin{titlepage}

\thispagestyle{empty}

\def\thefootnote{\fnsymbol{footnote}}

\begin{flushright}
IFT-UAM/CSIC-23-110
\end{flushright}

\vspace*{1cm}

\begin{center}

\begin{Large}
\textbf{
LHC Study of Third-Generation Scalar Leptoquarks \\[.25em] with Machine-Learned Likelihoods
}
\end{Large}

\vspace{1cm}

{\sc
Ernesto~Arganda$^{1, 2}$%
\footnote{{\tt \href{mailto:ernesto.arganda@uam.es}{ernesto.arganda@uam.es}}}%
, Daniel~A.~D\'{\i}az$^{2}$%
\footnote{{\tt \href{mailto:daniel.diaz@fisica.unlp.edu.ar}{daniel.diaz@fisica.unlp.edu.ar}}}%
, Andres D. Perez$^{1}$%
\footnote{\tt \href{mailto:andresd.perez@uam.es	}{andresd.perez@uam.es}}%
, \\ Rosa~M.~Sand\'a Seoane$^{1}$%
\footnote{{\tt \href{mailto:r.sanda@csic.es}{r.sanda@csic.es}}}%
and Alejandro~Szynkman$^{2}$%
\footnote{{\tt \href{mailto:szynkman@fisica.unlp.edu.ar}{szynkman@fisica.unlp.edu.ar}}}%

}

\vspace{0.5truecm}

{\sl
$^1$Departamento de Física Teórica and Instituto de F\'{\i}sica Te\'orica UAM-CSIC, \\
Universidad Autónoma de Madrid, Cantoblanco, 28049 Madrid, Spain

\vspace*{0.15cm}

$^2$IFLP, CONICET - Dpto. de F\'{\i}sica, Universidad Nacional de La Plata, \\ 
C.C. 67, 1900 La Plata, Argentina
}

\vspace*{2mm}

\end{center}

\vspace{0.1cm}

\renewcommand*{\thefootnote}{\arabic{footnote}}
\setcounter{footnote}{0}

\begin{abstract}
\noindent
We study the impact of machine-learning algorithms on LHC searches for leptoquarks in final states with hadronically decaying tau leptons, multiple $b$-jets, and large missing transverse momentum. Pair production of scalar leptoquarks with decays only into third-generation leptons and quarks is assumed. Thanks to the use of supervised learning tools with unbinned methods to handle the high-dimensional final states, we consider simple selection cuts which would possibly translate into an improvement in the exclusion limits at the 95$\%$ confidence level for leptoquark masses with different values of their branching fraction into charged leptons.  In particular, for intermediate branching fractions, we expect that the exclusion limits for leptoquark masses extend to $\sim$1.3 TeV. As a novelty in the implemented unbinned analysis, we include a simplified estimation of some systematic uncertainties with the aim of studying their possible impact on the stability of the results. Finally, we also present the projected sensitivity within this framework at 14 TeV for 300 and 3000 fb$^{-1}$ that extends the upper limits to $\sim$1.6 and $\sim$1.8 TeV, respectively. \github{https://github.com/AndresDanielPerez/LHC-3rd-gen-Leptoquarks-with-MLL} 
\end{abstract}

%\newpage

\end{titlepage}

\tableofcontents

\section{Introduction}
\label{intro}

Leptoquarks (LQs) are scalar or vector color-triplet bosons, with fractional electric charge and with both baryon and lepton numbers, that emerge from many different new physics models (for seminal papers, see for instance Refs.~\cite{Pati:1974yy,Georgi:1974sy,Georgi:1974yf,Fritzsch:1974nn,Pati:1975md,Dimopoulos:1979es,Dimopoulos:1979sp,Eichten:1979ah,Schrempp:1984nj,Wudka:1985ef,Buchmuller:1986iq,Buchmuller:1986zs,Angelopoulos:1986uq,Barbier:2004ez}) and can be searched for systematically at the LHC~\cite{Queiroz:2014pra,Diaz:2017lit,Bhaskar:2021gsy}. The most current motivation for considering this type of hypothetical particles is the fact that LQs which couple to third-generation leptons and quarks~\cite{Fajfer:2012jt,Tanaka:2012nw,Sakaki:2013bfa,Hiller:2014yaa, Gripaios:2014tna,Calibbi:2015kma,Freytsis:2015qca,Bhattacharya:2015ida,Bauer:2015knc,Fajfer:2015ycq,Dumont:2016xpj,DiLuzio:2017chi,Buttazzo:2017ixm,Kumar:2018kmr,Angelescu:2021lln,Marzocca:2021azj,Freitas:2022gqs,Goncalves:2023qpz,Banik:2023ogi,Florez:2023jdb} could provide an explanation for the $B$ anomalies observed in several measurements~\cite{Belle:2007qnm,Belle:2010tvu,BaBar:2012obs,BaBar:2013mob,LHCb:2014cxe,LHCb:2015gmp,Belle:2015qfa,Belle:2016xuo,LHCb:2016ykl,Belle:2016ure,Belle:2016dyj,Belle:2016fev,Belle:2017oht,CMS:2017rzx,LHCb:2017rln,LHCb:2017vlu,ATLAS:2018gqc,Belle:2019rba,LHCb:2020zud,LHCb:2020gog,LHCb:2021trn,Belle-II:2021rof,LHCb:2021zwz,LHCb:2022qnv,LHCb:2022vje,CMS:2022mgd,LHCb:2023zxo,Belle-II:2023esi,LHCb:2023gel,LHCb:2023gpo}.

Indeed, the search for LQs at the LHC represents a very intensive experimental program carried out by the ATLAS~\cite{ATLAS:2011atv,ATLAS:2011zhi,ATLAS:2012aq,ATLAS:2013oea,ATLAS:2015hsi,ATLAS:2016wab,ATLAS:2019ebv,ATLAS:2019qpq,ATLAS:2020dsf,ATLAS:2020dsk,ATLAS:2020xov,ATLAS:2021oiz,ATLAS:2021yij,ATLAS:2021mla,ATLAS:2021jyv,ATLAS:2022wcu,ATLAS:2023uox,ATLAS:2023kek,ATLAS:2023vxj,ATLAS:2023prb} and CMS~\cite{CMS:2010ssz,CMS:2010chx,CMS:2011zfm,CMS:2012iln,CMS:2012bfi,CMS:2012cyn,CMS:2014wpz,CMS:2015nep,CMS:2015xzc,CMS:2015gua,CMS:2016fxb,CMS:2017xcw,CMS:2018svy,CMS:2018qqq,CMS:2018txo,CMS:2018lab,CMS:2018oaj,CMS:2018iye,CMS:2018ncu,CMS:2018yiq,CMS:2020wzx,CMS:2021far,CMS:2022nty,CMS:2022goy,CMS:2022ncp,CMS:2023bdh,CMS:2023qdw} collaborations. More interestingly, many of these analyses already incorporate the application of modern machine learning (ML) techniques most commonly used in high-energy physics (for recent reviews, see for example Refs.~\cite{Larkoski:2017jix,Guest:2018yhq,Albertsson:2018maf,Radovic:2018dip,Carleo:2019ptp,Bourilkov:2019yoi,Feickert:2021ajf,Schwartz:2021ftp,Karagiorgi:2021ngt,Shanahan:2022ifi,Plehn:2022ftl}). Among these ML tools, the binned likelihood estimation through the use of the entire discriminant ML output is worth mentioning~\cite{ATLAS:2012byx,ATLAS:2015eiz,CMS:2017kxn,CMS:2018fdh,CMS:2018amb,CMS:2020bfa,ATLAS:2021kqb,ATLAS:2022ooq}. These methods are also being implemented in phenomenological analyses, as in the recent Refs.~\cite{Ghosh:2023ocz} and~\cite{Florez:2023jdb}, where ML studies on third-generation leptoquarks are presented. 

To date, none of these experimental analyses, even with sophisticated ML tools, has been able to find significant deviations from the standard model (SM) predictions, so it is legitimate to ask whether the use of unbinned methods would have the potential to improve on these binned methods. In order to try to answer this question, we work within the framework of Machine-Learned Likelihoods (MLL)~\cite{Arganda:2022qzy,Arganda:2022mrd,Arganda:2022zbs}, a method that combines supervised ML classification techniques with likelihood-based inference tests, allowing to estimate the experimental sensitivity of high-dimensional data sets without the need of binning the score output to extract the resulting one-dimensional signal and background probability density functions (PDFs), by means of the use of kernel density estimators (KDE)~\cite{RosenblattKDE,ParzenKDE}. It is well known that the major drawback of unbinned methods is the lack of knowledge about how to properly introduce nuisance parameters in the likelihood estimation, which makes them unsuitable for experimental analyses at present. As an initial approach to this issue and mainly as a test of the stability of our results under a simplified addition of some systematic uncertainties, we develop for the first time within the MLL framework a strategy to include them in the unbinned analysis and to estimate their impact on the calculation of signal significances.

In this article we apply the MLL framework to the study of the double production at the LHC of up-type and down-type scalar LQs exclusively coupled to third-generation SM fermions. LQs only coupled to the third generation are interesting in their own right and they have been extensively studied from both experimental and phenomenological perspectives (see for instance~\cite{Bhaskar:2021gsy,ATLAS:2021oiz,ATLAS:2023uox,CMS:2020wzx,CMS:2023qdw,Ghosh:2023ocz})~\footnote{Certainly, in order to address the issue of $B$ anomalies commented at the beginning of this section, LQs with non-diagonal flavor couplings are necessary.}. For this scenario we consider the decays of up-type and down-type LQs into $t \nu_{\tau}/b \tau$ and $b \nu_{\tau}/t \tau$ final states, respectively, and compute the expected exclusion limits in the [BR$(LQ_3^{u/d}\rightarrow q l)$, $m(LQ_3^{u/d})$] plane, where  BR$(LQ_3^{u/d}\rightarrow q l)$ is the branching fraction of a given channel to a quark and a tau lepton and $m(LQ_3^{u/d})$ the mass of the LQ. It is worth stressing that the study of third-generation LQs is not related to a limitation of the MLL framework. In fact, we expect this to be suitable and even promising for the study of LQs coupled to other or mixed generations, although dedicated searches for specific signals and backgrounds, beyond the scope of this work, would be necessary in order to asses a quantitative impact.

The paper is structured in the following way. In Section~\ref{ph-frame} we present the main features of the LQ model we work with and the LHC experimental setup in which we perform our phenomenological analysis. In Section~\ref{simulation} we describe the signal and background simulation sampling and our event selection criteria. Section~\ref{unbinned-method} is devoted to our collider analysis with the unbinned method MLL. Our results are shown in Section~\ref{results}, in which we have incorporated in the MLL method an approach to the inclusion of systematic uncertainties, considering only those that directly affect the most relevant variables with which we feed the ML classifier, individually and without correlations. We also show the prospects for the LHC with a center-of-mass energy of 14 TeV. Finally, our manuscript ends in Section~\ref{conclu} where we discuss our main conclusions. 

\section{Phenomenological Framework}
\label{ph-frame}

Leptoquarks naturally arise from many extensions of the SM, such as Grand Unification Theories (GUTs), technicolor scenarios, or composite models as new hypothetical fields. LQs are either scalar or vector fields and can couple to quark-lepton currents, therefore they involve local interactions between quarks and leptons, being this feature their main signature. Following Ref.~\cite{Buchmuller:1986zs}, the effective theory containing the most general couplings, invariant under $SU(3)\times SU(2) \times U(1)$, for scalar leptoquarks which couple third-generation quark to lepton fields preserving baryon and lepton number conservation is defined by the Lagrangian,

\begin{align}
\mathcal{L} &=   \left( g_{1L}\bar{q}^c_L i \tau_2 l_L+ g_{1R}\bar{t}^c_R \tau_R  \right) S_1+ \tilde{g}_{1R}\bar{b}^c_R \tau_R \tilde{S}_1 +g_{3L}\bar{q}^c_L i \tau_2 (\bm{\tau} \cdot \bm{S}_3) l_L \nonumber \\
&\quad +\left(h_{2L}\bar{t}_R l_L+h_{2R}\bar{q}_L i \tau_2 \tau_R \right) R_2 + \tilde{h}_{2L}\bar{b}_R l_L \tilde{R}_2 + c.c., 
\label{BRW_lagrangian}
\end{align}
where $q_L$ and $l_L$ denote left-handed quark and lepton doublets, and $\tau_R$, $b_R$, and $t_R$ are $SU(2)$ singlets for right-handed tau leptons, and bottom and top quarks, respectively. Besides, $S_1$, $\tilde{S}_1$, $S_3$,  $R_2$, and $\tilde{R}_2$ stand for the scalar LQ fields in the interaction basis. Switching into the physical basis, we define up-type and down-type LQs, $LQ_3^u$ and $LQ_3^d$, as the lighter linear combinations of $\tilde{R}_2$, $R_2$ and $S_3^*$, and $S_1^*$, $S_3^*$ and $\tilde{R}_2$, respectively. In this basis, $LQ_3^u$($LQ_3^d$) has electrical charges $\frac{2}{3}$($-\frac{1}{3})e$.\\

In Ref.~\cite{ATLAS:2021jyv} the ATLAS Collaboration presented a search for new phenomena at the LHC in processes involving tau leptons, $b$-jets, and missing transverse momentum in the final state. The analyzed data set corresponds to a center-of-mass energy of $\sqrt{s}=13$ TeV for proton-proton collisions and an integrated luminosity of $139$ fb$^{-1}$. The results were interpreted within two simplified benchmark model scenarios, one of which considers the pair production of up-type or down-type scalar leptoquarks. In both cases, the LQs were assumed to couple only to third-generation SM fermions following the minimal Buchmüller-Rückl-Wyler model yielding the decays $LQ_3^u\rightarrow t \nu_{\tau}/b \tau$ for up-type LQs, and $LQ_3^d\rightarrow b \nu_{\tau}/t \tau$ for down-type LQs. The model parameters consisted of the masses of the LQs, $m(LQ_3^{u/d})$, and their respective branching fractions for decays into a quark and a charged lepton, $\beta=$ BR$(LQ_3^{u/d}\rightarrow q l)$. 
%The ATLAS collaboration finds that scalar LQs with masses up to $1.25$ TeV are excluded at $95\%$ $CL$.\\

We are interested here in the single-tau signal region reported by the ATLAS collaboration, which corresponds to the analysis involving the LQ simplified model. In particular, in order to explore the impact on the exclusion limits in the [BR$(LQ_3^{u/d}\rightarrow q l)$, $m(LQ_3^{u/d})$] plane, we focus on the multi-bin signal region which corresponds to the one where those limits were obtained by ATLAS. This signal region is characterized by selection requirements on the missing transverse momentum ($E_T^{\text{miss}}$), the tau-lepton transverse momentum ($p_T(\tau)$), the sum of the transverse momenta corresponding to the tau lepton and the two leading jets ($s_T$), the tau-lepton transverse mass ($m_T(\tau)$), and the sum of the transverse masses of the $b$-jets, ($\sum m_T(b_1,b_2)$). In the next section, we will discuss the details concerning these requirements as well as the specifics and simulation of the backgrounds. Besides, we will also elaborate on the simulation of the signal data sets corresponding to LQ masses ranging from $800$ GeV up to $1800$ GeV and a fixed value of BR$(LQ_3^{u/d}\rightarrow q l)=0.5$ that encompasses different values of the LQ couplings. We will also explain how we extended the analysis to other branching fractions.\\ 

\section{Event Simulation and Selection}
\label{simulation}

The proposed signal under study is the double production of up-type or down-type leptoquarks at the LHC. In the up-type case, one of the letpoquarks decays into a top quark and a neutrino and the other one into a bottom quark and a $\tau$ lepton:
\begin{equation}
pp \to LQ_3^{u} LQ_3^{u} \to t\nu + b\tau \,.
\end{equation}
In the down-type case, one of the letpoquarks decays into a bottom quark and a neutrino and the other one into a top quark and a $\tau$ lepton:
\begin{equation}
pp \to LQ_3^{d} LQ_3^{d} \to b\nu + t\tau \,.
\end{equation}
The simulated sample for leptoquarks was generated with {\sc MadGraph5\_aMC@NLO} \cite{Alwall:2014hca} at leading order in QCD with the NNPDF2.3 LO PDF set \cite{Ball:2012cx}, and at a center-of-mass energy of 13 TeV. The UFO model is the one developed in~\cite{Mandal:2015lca}.  Events were processed with {\sc Pythia} \cite{Sjostrand:2014zea,Sjostrand:2007gs} for parton showering and hadronization, and {\sc Delphes} \cite{deFavereau:2013fsa} for fast detector simulation, using the default ATLAS card. 

In order to reproduce the conditions from search~\cite{ATLAS:2021jyv} we considered the same object identification criteria defined by the ATLAS collaboration, summarized in Table~\ref{table:cuts}. For validating our event generation pipeline, we reproduced the $p_T(\tau)$ distribution in~\cite{ATLAS:2021jyv}  for a $m(LQ_3^{u})=1.2$ TeV and $\beta=0.5$, for the single-tau final state. This final state is defined by the presence of exactly one $\tau_\text{had}$ and at least two $b$-jets ($n_b\geq2$), with a light-lepton veto ($e/\mu$) and a lower bound on $E_T^\text{miss}$ at 280 GeV. There is also a lower bound on the sum of the transverse masses of the $b$-jets,
\begin{equation}
    \sum m_T(b_1,b_2)=m_T(b_1)+m_T(b_2),
\end{equation}
where $b_1$ and $b_2$ are the two leading $b$-tagged jets and $m_T$ the transverse mass defined as,
\begin{equation}
    m_T(A)\equiv m_T(\mathbf{p}_T (A), \mathbf{E}_T^\text{miss})=[2 p_T(A) E_T^\text{miss} (1-\cos \Delta \phi(\mathbf{p}_T (A), \mathbf{E}_T^\text{miss}))]^{1/2}.
\end{equation}
The aforementioned $p_T(\tau)$ distribution uses the multi-bin signal region, defined by bounds on the $\tau$ transverse momentum as well as its transverse mass and $s_T$, being the latter defined by
\begin{equation}
    s_T=p_T(\tau_\text{had})+p_T(j_1)+p_T(j_2),
\end{equation}
where $j_1$ and $j_2$ are the two leading jets. We fairly reproduced the $p_T(\tau)$ shape distribution, and fit the results with an additional global normalization factor.
The bounds on these variables are shown in Table~\ref{table:cuts_2}.

Background events were simulated at LO in QCD with {\sc MadGraph5\_aMC@NLO}, and subsequently processed with {\sc Pythia} and {\sc Delphes}. We considered the main contributions in the multi-bin single-tau signal region: $t\Bar{t}$ (with 1 real $\tau_{had}$), fake-$t\Bar{t}$ (with no real $\tau_{had}$), single-top, $W+$jets, %Z$+$jets, 
$t\Bar{t}H$ and $t\Bar{t}V$ (with $V=W+Z$)~\footnote{The $Z+$jets background was not considered as it turned out to be subleading for the selection cuts employed in this work.}.

It is important to mention here that, as simplified assumptions of the work, in the analysis no QCD corrections are taken into account in the signal and background processes. The ATLAS study~\cite{ATLAS:2021jyv}, on the other hand, considers two extra jets in the signal simulation, while backgrounds are generated with a POWHEG box~\cite{Nason:2004rx,Frixione:2007nw,Frixione:2007vw,Alioli:2010xd}.

\begin{table}
\centering
\begin{tabular}{ cc } 
 \hline
 Object \hspace{0.5cm} & \hspace{0.5cm} Identification Criteria \\ 
  \hline
hadronically-decaying tau \hspace{0.5cm} & \hspace{0.5cm} $p_T> 20$ GeV;\hspace{0.5cm} $|\eta|<2.5$; \hspace{0.25cm}
$|\eta| \not\in  (1.37,1.52)$ 
 \hspace{0.25cm}   \\

jets \hspace{0.5cm} & \hspace{0.5cm} $p_T> 20$ GeV;\hspace{0.5cm} $|\eta|<2.8$   \\
b-tagged jets \hspace{0.5cm} & \hspace{0.5cm} $p_T> 20$ GeV;\hspace{0.5cm} $|\eta|<2.5$   \\
electrons \hspace{0.5cm} & \hspace{0.7cm} $p_T> 10$ GeV;\hspace{0.5cm} $|\eta|<2.47$   \\
muons \hspace{0.5cm} & \hspace{0.5cm} $p_T> 10$ GeV;\hspace{0.5cm} $|\eta|<2.7$   \\

 \hline
\end{tabular}
 \caption{Object identification criteria employed in this work.}
\label{table:cuts}
\end{table}

\begin{table}
\centering
\begin{tabular}{ cc } 
 \hline
 ATLAS cuts \hspace{0.5cm} & \hspace{0.5cm} Loose cuts \\ 
  \hline
 \multicolumn{2}{c}{single-tau final state, $n_{\tau_\text{had}}$=1} \\ 
 \multicolumn{2}{c}{at least 2 $b$-jets, $n_b\geq2$} \\ 
 \multicolumn{2}{c}{no light leptons} \\ \multicolumn{2}{c}{$E_T^\text{miss}>280$ GeV} \\ 
 $p_T(\tau)> 50$ GeV \hspace{0.5cm} & \hspace{0.5cm} $p_T(\tau)> 20$ GeV \\ 
  $\sum m_T(b_1,b_2)>700$ GeV \hspace{0.5cm} & \hspace{0.5cm} - \\ 
  $m_T(\tau_\text{had})>150$ GeV \hspace{0.5cm} & \hspace{0.5cm} - \\
  $s_T>600$ GeV \hspace{0.5cm} & \hspace{0.5cm} - \\
 \hline
\end{tabular}
 \caption{Selection cuts considered by ATLAS Collaboration~\cite{ATLAS:2021jyv} used to validate our pipeline, and loosened cuts employed in our multivariate analysis.}
\label{table:cuts_2}
\end{table}

\begin{figure}
  \centering
  \includegraphics[width=0.8\textwidth]{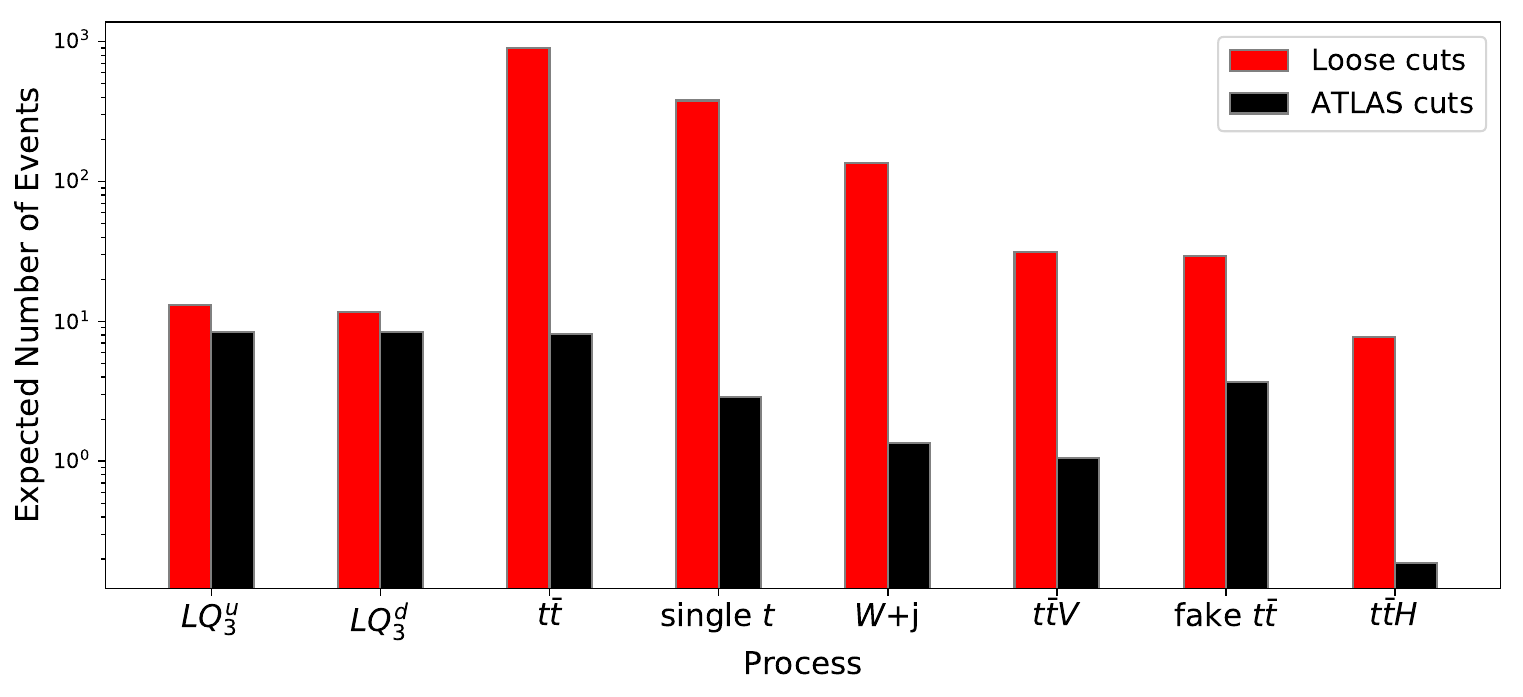}
\caption{Expected number of events at the LHC $\sqrt{s}=13$ TeV and 139 fb$^{-1}$ for both types of LQs with $m(LQ_3^{u/d})=1.2$ TeV and $\beta=0.5$, and the main backgrounds considered in this work.  Two selection cuts are shown, in red the ones used throughout this work and in black the ones described by ATLAS~\cite{ATLAS:2021jyv}.}
\label{LQ-num-events}
\end{figure}

For our multivariate analysis, event selection criteria are loosened to fully exploit the discrimination power of machine-learning classifiers. We used the same criteria for jets, $b$-jets, and lepton identification. We keep on working with the single-tau final state ($n_{\tau_\text{had}}$=1, $n_b\geq2$, no light leptons, and  $E_T^\text{miss}>280$ GeV), but without any requirement in high-level observables $m_T(b_1,b_2)$, $m_T(\tau_\text{had})$ and $s_T$. Also the lower bound on $p_T(\tau)$ was relaxed to 20 GeV. In Table~\ref{table:cuts_2} we show both selection strategies. These loose cuts also ease data simulation, since ATLAS cuts for the single-tau multi-bin region are very tight and allow to retain only a few Monte Carlo events per simulation: we obtain that a fraction of $\sim 10^{-4}$ background events survives the ATLAS cuts with respect to the generated ones, while considering the loose cuts the surviving fraction is 2 orders of magnitude larger. For the signal events the impact is milder, from $\sim 10^{-1}-10^{-2}$ to $\sim 10^{-1}$, since the ATLAS cuts are designed to define a signal enriched region. In Fig.~\ref{LQ-num-events} we present the expected number of events for both selection criteria at the LHC $\sqrt{s}=13$ TeV and 139 fb$^{-1}$. For signals, we have chosen as benchmark $m(LQ_3^{u/d})=1.2$ TeV and $\beta=0.5$. The backgrounds are in descending order of relevance considering the loose cuts and we can see that the hierarchy of the main backgrounds is modified with respect to the ATLAS cuts. Another important characteristic is that the signal-to-background ratio decreases significantly, however, this is intentional in order to highlight that for the multivariate analysis, one does not need to design very tight signal regions which could inadvertently remove significant information that would otherwise help in the discrimination task.

We simulated events on the mass range $m(LQ_3^{u/d})\in [800,1800]$ GeV, selecting benchmark points with a step of $200$ GeV in mass, and a fixed value of $\beta=0.5$. For each benchmark point, we simulate enough events such that we end up with $\sim$500k events after the selection cuts described above. Similarly, we have simulated enough background events to obtain a $\sim$500k data sample at detector level, considering the relative weight of each background channel.

Since the signal samples were generated with $\beta=0.5$, both leptoquark decay channels, either into a quark and a neutrino or into a quark and a charged lepton, are possible. These events can be reweighted to different branching fractions to derive limits in the $m(LQ_3^{u/d})$ vs BR$(LQ_3^{u/d} \rightarrow q \ell)$ parameter space. To this end, for every $m(LQ_3^{u/d})$ value we simulated a small sample of events, $\sim$50k, within the range $\beta \in [0, 1]$ to reweight the $\beta =0.5$ data according to their relative cross sections after selection cuts.

\section{Analysis Strategy with Unbinned Machine-Learned Likelihoods}
\label{unbinned-method}

\begin{figure}
  \centering
  \includegraphics[width=0.32\textwidth]{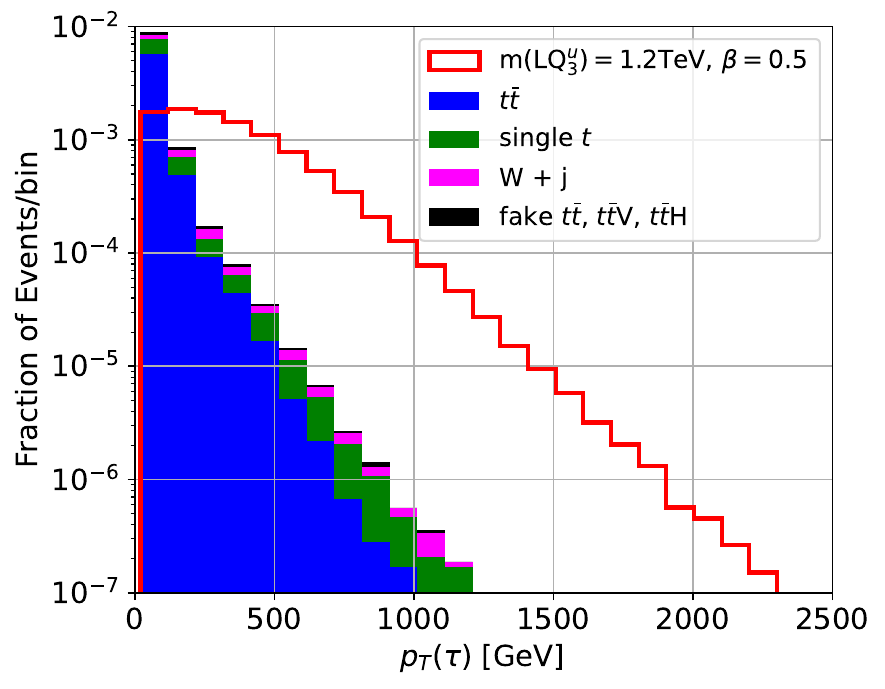}
  \includegraphics[width=0.32\textwidth]{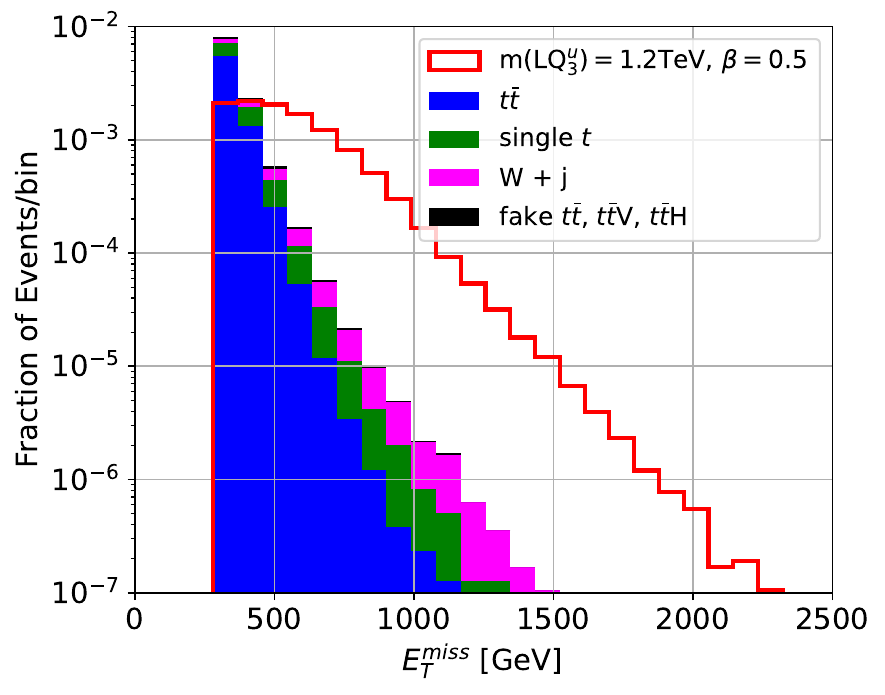}
  \includegraphics[width=0.32\textwidth]{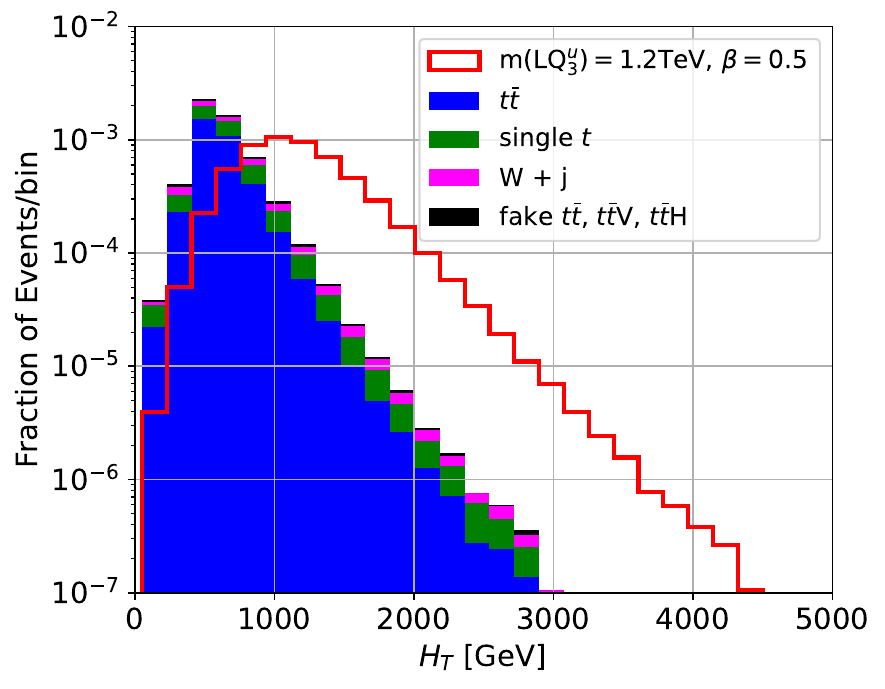}
\caption{Distributions of the most relevant variables for the ML discrimination: $p_T(\tau)$ (left panel), $E_T^\text{miss}$ (middle panel), and $H_T$ (right panel). The signal distributions (red curve) correspond to a $LQ^u_3$ with mass equal to 1.2 TeV and $\beta=0.5$ as a benchmark. The stacked histograms show the SM background contributions. Minor backgrounds, i.e. fake-$t\bar{t}$, $t\bar{t}V$, and $t\bar{t}H$, are grouped together.}
\label{distributions}
\end{figure}

A large but simple set of discriminating variables is used to feed the ML algorithm: $p_T$, $\eta$ and $\phi$ of the reconstructed $\tau_{had}$, $b_1$ and $b_2$ (the two leading $b$-tagged jets); the missing transverse momentum information, $E_T^\text{miss}$ and $\phi(E_T^\text{miss})$; the number of identified jets, $n_\text{jets}$, the number of $b$-tagged jets, $n_b$, and the hadronic activity $H_T=\sum_{i=1}^{n_\text{jets}} p_T^{j_i}$. Following the same motivation explained for the `loose' event selection criteria, notice that we are not considering any high-level observable. As an example, in Fig.~\ref{distributions} we present the signal ($m(LQ^u_3)=1.2$ TeV and $\beta=0.5$) and background distributions of the most relevant variables for the ML discrimination, $p_T(\tau)$, $E_T^\text{miss}$, and $H_T$, as we will see below (the distributions for $LQ^d_3$ are similar).

\begin{figure}
  \centering
  \includegraphics[width=0.49\textwidth]{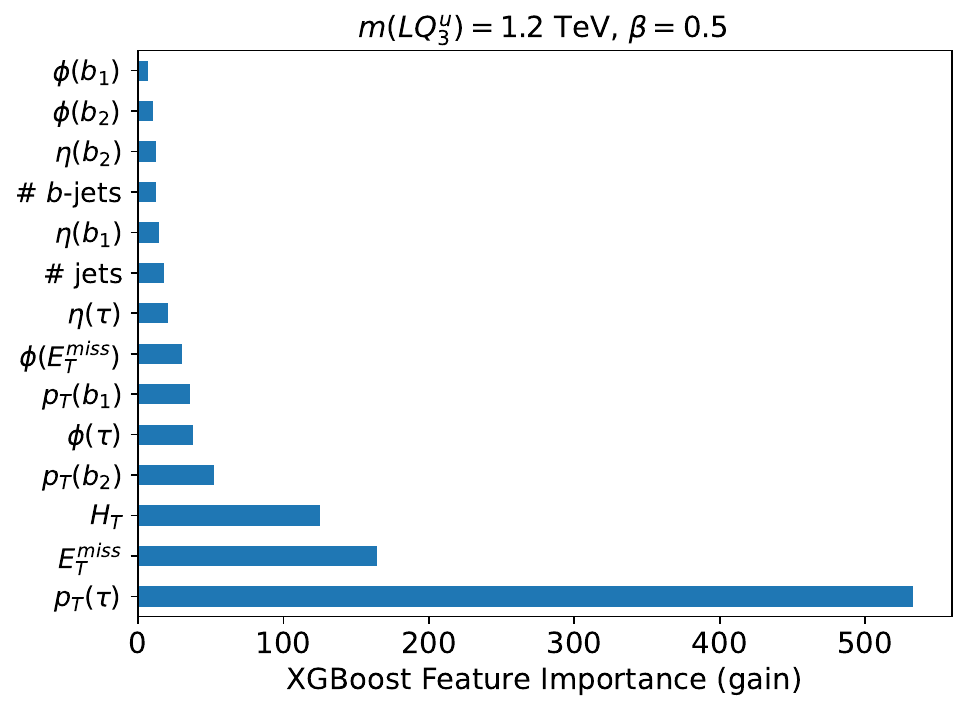}
  \includegraphics[width=0.49\textwidth]{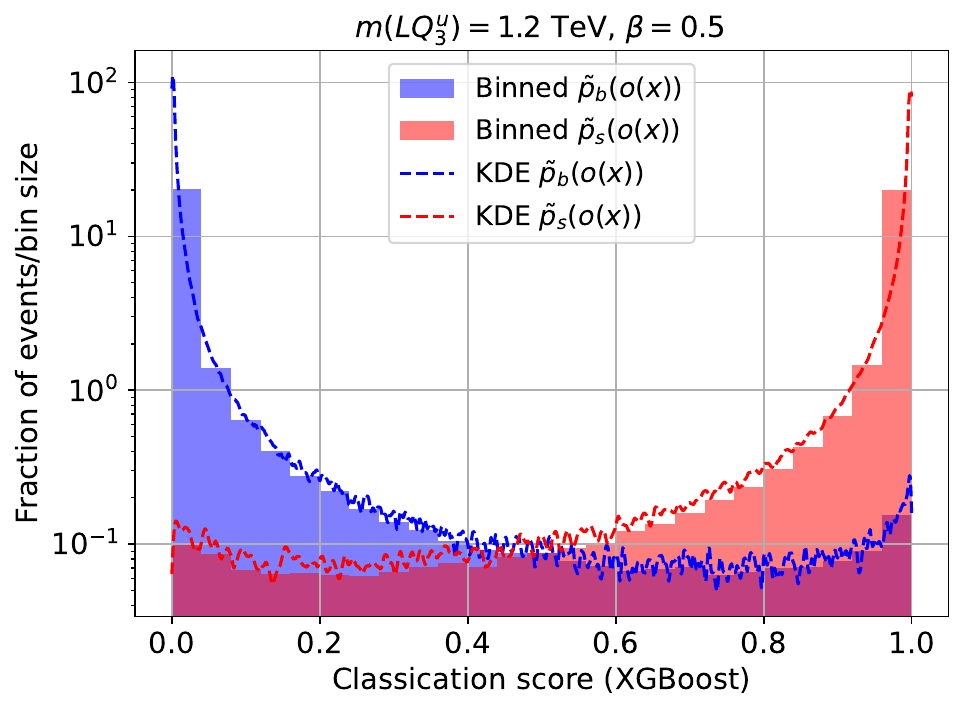}
\caption{Left panel: feature importance score (gain metric), which indicates how useful or valuable each feature was for the ML discrimination. Right panel: output of the {\tt XGBoost} classifier when tested with only pure background (blue) or pure signal (red) samples. The dashed curves correspond to the PDFs obtained with the KDE method. Both panels consider as signal a $m(LQ^u_3)=1.2$ TeV and $\beta=0.5$ as a benchmark.}
\label{XGBoost-outputs}
\end{figure}

For each value of $m(LQ_3^{u/d})$, we trained a supervised per-event classifier, using the {\tt XGBoost} toolkit~\cite{Chen:2016btl,Chen:2016:XST:2939672.2939785}, with 500k events per class (balanced data set), as a binary classifier to distinguish signal ($S$) from background ($B$). In the background sample, we consider the relative weight of each background channel by its relative contribution after applying our selection cuts. For further details about the algorithm employed, the code is available at~\cite{MLL-LQ-code}. In the left panel of Fig.~\ref{XGBoost-outputs} we present the feature importance score, considering the gain metric that measures the relative contribution of the corresponding feature. A higher value of this metric when compared to another feature implies it is more important for generating a prediction. We employ the same data set as in Fig.~\ref{distributions}, and as anticipated before, $p_T(\tau)$, $E_T^\text{miss}$, and $H_T$ are the most relevant features. For different values of $m(LQ^{u/d}_3)$ there are small changes in the hierarchy, although the general trend is not modified.

The output of the ML classifier, $o(x)$, for this example is shown on the right panel of Fig.~\ref{XGBoost-outputs} when tested with only pure background or pure signal new samples, blue and red histograms, respectively. The output, $o(x) \in [0,1]$, quantifies if an event is either signal-like (near 1) or background-like (near 0). Within our setup, we obtained an area under the receiver operating characteristic (ROC) curve of 0.992. 

To estimate the exclusion limits in this work, we will exploit the entire discriminant ML output by comparing a binned and an unbinned approach. It is known that the binning process has an inherent drawback associated to the loss of information of the probability densities of each event inside each bin, which in turn impacts in the likelihood estimation, potentially decreasing the significance. This loss of information can be reduced by making the bin width small enough, but this solution is usually limited by the finite statistics, that renders the binned likelihood density estimation unreliable for large number of bins. On the other hand, unbinned methods not only preserve the granularity of individual data points, potentially offering a more accurate representation, but also allow greater flexibility in capturing complex distributions and subtle patterns in the data because they do not average information across bins.

For the histogram-based approach, the traditional Binned Likelihood (BL) method~\cite{Cowan:2010js} is used, where a likelihood function is built as the product of Poisson probability functions. Then, the full ML output is binned to find the expected number of signal and background events in each bin (see the histograms in the right panel of Fig.~\ref{XGBoost-outputs}).

For the unbinned approach, the strategy applied in this work is based on the MLL framework~\footnote{For more details, we refer the reader to the original MLL articles~\cite{Arganda:2022qzy,Arganda:2022mrd,Arganda:2022zbs} where the method is developed.}, that can be schematically summarized as follows:
\begin{itemize}
\item It uses a binary classifier to discriminate between signal and background (in this case {\tt XGBoost}), which is fed with event-by-event variables. This allows us to convert a high-dimensional problem into a single-dimensional one, based on the score of the classifier output.
\item From the entire unbinned classifier output, we estimate the signal and background PDFs, $\tilde{p}_{s}(o(x))$ and $\tilde{p}_{b}(o(x))$, using KDE to fit the classifier output when tested with only pure background or pure signal samples, respectively (see the dashed curves in the right panel of Fig.~\ref{XGBoost-outputs}). 
\item Knowing the signal and background PDFs, we compute the likelihood function of the hypothesis tests of interest. MLL has both discovery and exclusion tests included, which allows us to estimate both the signal significance of discovery (5$\sigma$) and evidence (3$\sigma$) and also to impose exclusion limits at 95\% confidence level (CL, 1.64$\sigma$)~\cite{Cowan:2010js}.
\end{itemize}

Even though the KDE method is called a non-parametric method for density estimation because it does not assume a specific functional form for the underlying distribution, it involves one parameter known as ``bandwidth''. This parameter controls the degree of smoothness of the estimated density function, and can be chosen or estimated from the data. A larger bandwidth leads to a smoother estimate, while a smaller bandwidth results in a more variable estimate that is sensitive to local fluctuations. Throughout our work, we selected the bandwidth parameters individually for the signal and background PDFs with a grid search using the \texttt{GridSearchCV} function available inside the \texttt{sklearn.model\textunderscore selection}~\cite{scikit-learn} python package, which gives as an output the bandwidth which maximizes the data likelihood in a 5-fold cross-validation strategy. For further details about the implementation of the unbinned method and the KDE algorithm see~\cite{MLL-LQ-code}, where the code for this work is available.

\section{Results}
\label{results}

Assuming there is no significant excess, we use the exclusion hypothesis test and compute the exclusion limits for each point in the parameter space of the scalar LQ mass, $m(LQ_3^{u/d})$, and its branching fraction into a quark and a charged lepton, $\beta$ = BR$(LQ_3^{u/d} \rightarrow q \ell)$. Finally, we define the exclusion limits at 95\% CL as the curve where the significance is equal to 1.64$\sigma$. 

\begin{figure}
  \centering
  \includegraphics[width=0.49\textwidth]{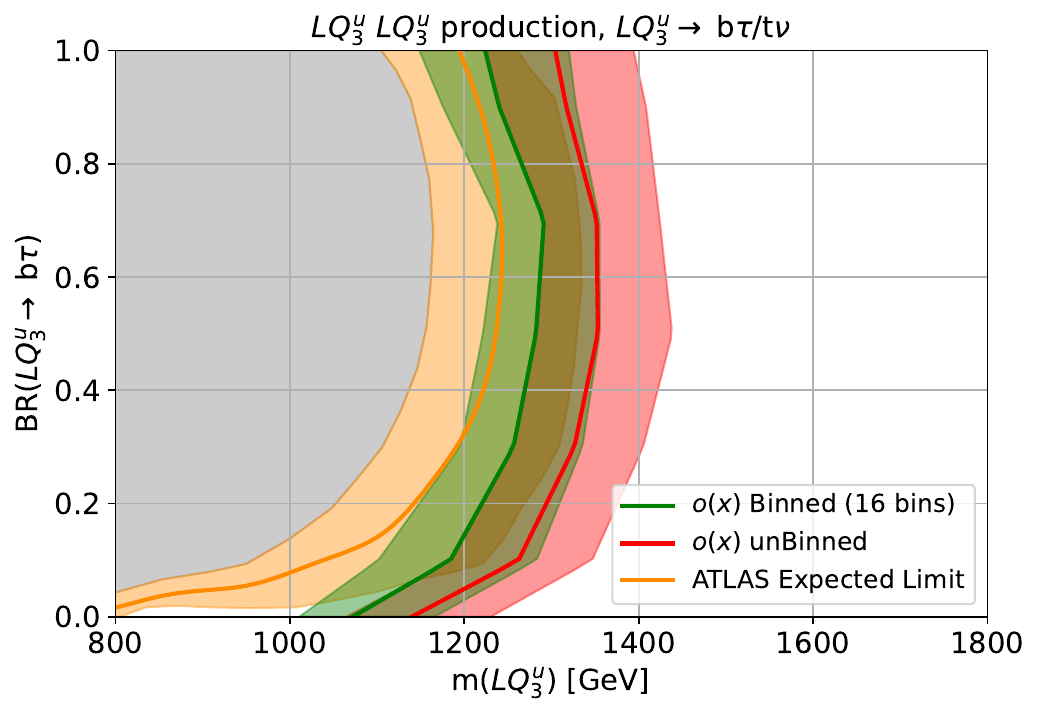}
  \includegraphics[width=0.49\textwidth]{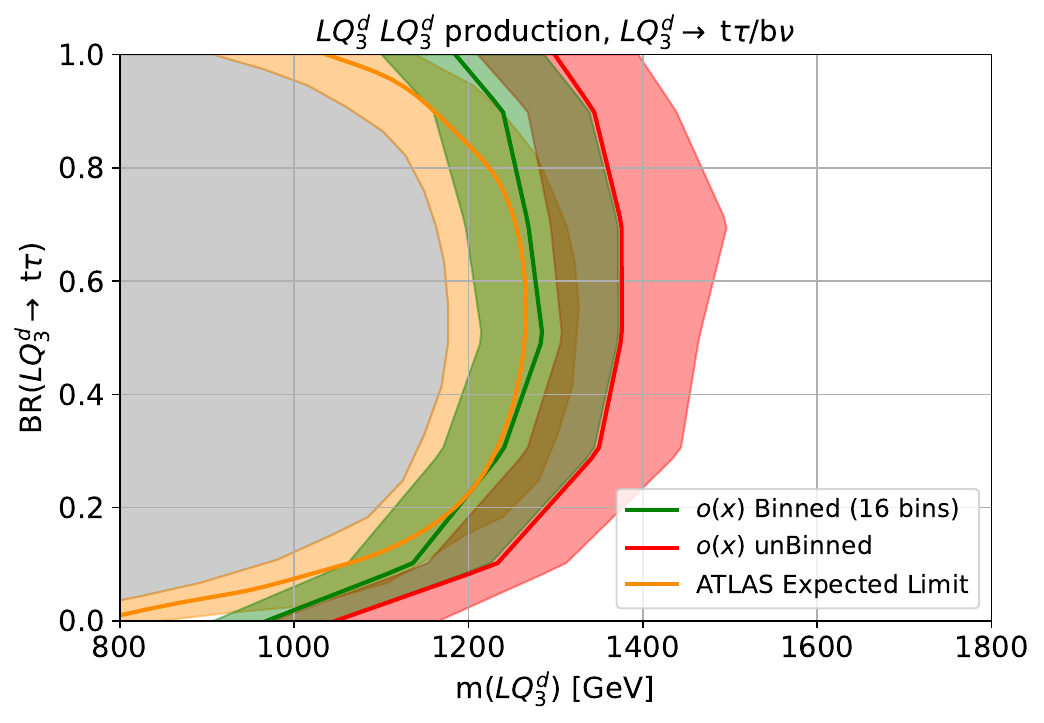}
\caption{Expected exclusion contours at the 95\% CL for the third-generation up-type (left panel) and down-type (right panel) scalar-leptoquarks, in the leptoquark mass, $m(LQ_3^{u/d})$, and branching fraction into a quark and a charged lepton, BR$(LQ_3^{u/d} \rightarrow q \ell)$, space. The limits are derived using a binned likelihood method and an unbinned approach, both considering the entire output of the ML classifier, $o(x)$. The colored regions represent $\pm 1\sigma_{stat}$, but no systematic uncertainties are included. As a reference, we present the expected exclusion-limit contours obtained by ATLAS in~\cite{ATLAS:2021jyv}.}
\label{LQ-nosys}
\end{figure}

In Fig.~\ref{LQ-nosys} we show the expected exclusion limits using a binned likelihood method (green) and an unbinned approach (red), both considering the full output of the ML classifier, $o(x)$. The colored regions show the impact of the statistical uncertainty ($\pm 1\sigma_{stat}$)~\footnote{To compute the significance, for each parameter point we generate 2k pseudo-experiments, then the statistical uncertainty is the dispersion of that sample calculated as the square root of its variance. We have also checked that increasing the number of pseudo-experiments up to 200k have negligible impact on the results.}. However, no systematic uncertainties were included in this calculation. As a reference, we also present the expected exclusion-limit contours obtained by ATLAS~\cite{ATLAS:2021jyv} through a binned likelihood test (\textit{multi-bin signal region}) considering high-level physical-based variables and the stringent selection cuts in Table~\ref{table:cuts_2} instead of using the ML output. In our implementation of the binned likelihood method we chose to work with 16 bins as in the search performed by ATLAS, which turns out to be very close to the maximum number of equal-sized bins that can be allowed when requiring at least 5 background events per bin, a common practice to ensure statistically robust and reliable results at LHC experiments. We have checked nonetheless that the results with the binned likelihood method do not change significantly when increasing the number of bins up to this maximum value.

For both types of leptoquarks, the expected exclusion contours extend to masses $\sim$1.28 TeV and $\sim$1.36 TeV for the binned and unbinned methods, respectively, and for intermediate values of BR$(LQ_3^{u/d} \rightarrow q \ell)$. Since the fraction of events with exactly one tau lepton decreases when BR$(LQ_3^{u/d} \rightarrow q \ell) \rightarrow 0$ or 1, the signal acceptance decreases which leads to lower mass values excluded. We can see that the multivariate analysis shows a tendency towards a possible improvement of the exclusion limits set by the ATLAS collaboration. Moreover, the unbinned method provides more stringent expected exclusion limits in the entire parameter space, although it is computationally more expensive.

We want to highlight that we have performed a series of cross-checks to assess the robustness of our procedure, and our results do not change significantly. Instead of working with a single ML output, $o(x)$, we averaged over the output of ten independent ML realizations, defined as $\langle o(x) \rangle = \frac{1}{10}\sum_{i}^{10}o_{i}(x)$, where each ML was trained with an independent data set. Consequently, the estimation of $\tilde{p}_{s,b}(o(x))$ using the KDE over the average variable $\langle o(x) \rangle$ turns out to be slightly smoother than the estimation over a single machine learning output $o(x)$ (shown for example on the right panel of Fig.~\ref{XGBoost-outputs}).

We also have checked that the numerical instabilities introduced by the bandwidth parameter do not affect our results. The influence of the bandwidth parameter is expected to be less pronounced for large data sets because the density estimate tends to converge to the true underlying distribution as the sample size increases. We employ 50k events to estimate the probability density functions with KDE using data cross-validation, and we have checked that increasing or decreasing the bandwidth by a factor $\sim 1-10$ with respect to that value does not significantly modify the results. The same conclusion holds if we fine tune the bandwidth search (with more computational cost) by increasing the sample size and/or decreasing the step size used by the search algorithm to determine the bandwidth.

Finally, with new and independent training/testing data sets we have cross-validated that the results using these samples are compatible within statistical fluctuations in both binned and unbinned methods.

\subsection{Approach to the Inclusion of Systematic Uncertainties}
\label{systematics}

This subsection aims to estimate the impact of some systematic uncertainties on our calculation of the expected exclusion limits. The procedure detailed below attempts to be a first approximation to evaluate the stability of our results, especially for the unbinned method, and in a multivariate-based approach, deals directly with the relevant kinematic variables used to feed the ML classifier instead of dealing with the underlying nuisance parameter affecting them.

First, we need to translate the systematic uncertainties in the physical-based space to an uncertainty of the ML classifier output space. We consider only uncertainties that affect directly the features used to train our ML algorithm and take the correlations among them not significant. Since the most relevant variables for the ML discrimination are $p_T$ of tau leptons, $\MET$, and $H_T$ (see the left panel of Fig.~\ref{XGBoost-outputs}), we consider systematic shifts of 5-10\%~\cite{ATLAS:2017mpa} on each of those variables individually. Then, inspired by~\cite{Arratia:2021otl,CMS:2022ytw}, the impact on the ML output is assessed by using the same test data set as in the original setup, with the whole set of events with all the kinematic variables used as input variables unchanged, except for the one that we choose to shift by a $5-10\%$. Then, we increase or decrease only the value of the selected variable in all the events of the data set by the same percentage. For example, to estimate the impact of $\Delta p_T(\tau)$, the systematic uncertainty of the tau transverse momentum, we take the ML algorithm trained with no uncertainties and evaluate it with two new test samples with all the variables unchanged but replacing $p_T(\tau) \rightarrow$ $p_T(\tau) \pm \Delta p_T(\tau)$, where $\Delta p_T(\tau)/p_T(\tau) = 0.1$. With this procedure, we obtain two ML outputs: $o(x)^{\pm}$, respectively. 

For the binned method, the uncertainty of the ML output $o(x_d)$ in each bin $d$ would be $\Delta o(x_d) = |o(x_d)^{+} - o(x_d)^{-}|$, and we can estimate the significance introducing it into the profile likelihood formulae~\cite{Cowan:error}. For the unbinned case, we do not have an expression to compute the significance including systematic uncertainties. Nevertheless, we can estimate its impact by repeating the entire unbinned procedure twice with $o(x)^{\pm}$. To be conservative, we take as the final result the outcome that provides the less restrictive limit, taking into account individually the results for each possible shift and in each of the three considered variables ($p_T(\tau)$, $\MET$, and $H_T$ ). The variable that most affects the results is $p_T(\tau)$.

\begin{figure}
  \centering
  \includegraphics[width=0.49\textwidth]{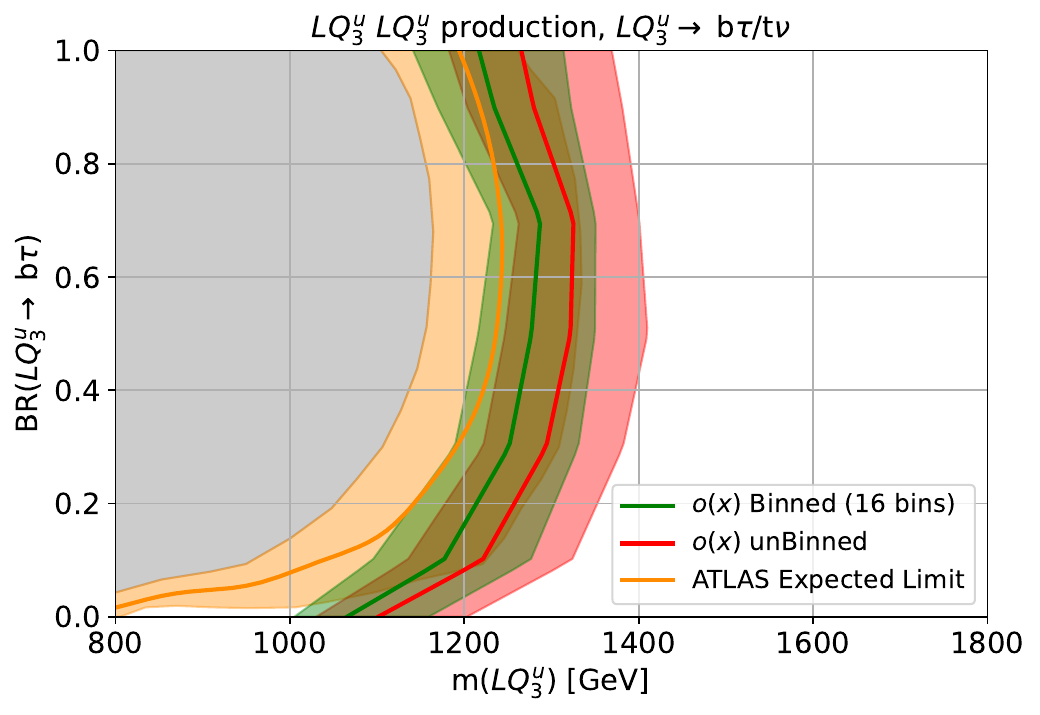}
  \includegraphics[width=0.49\textwidth]{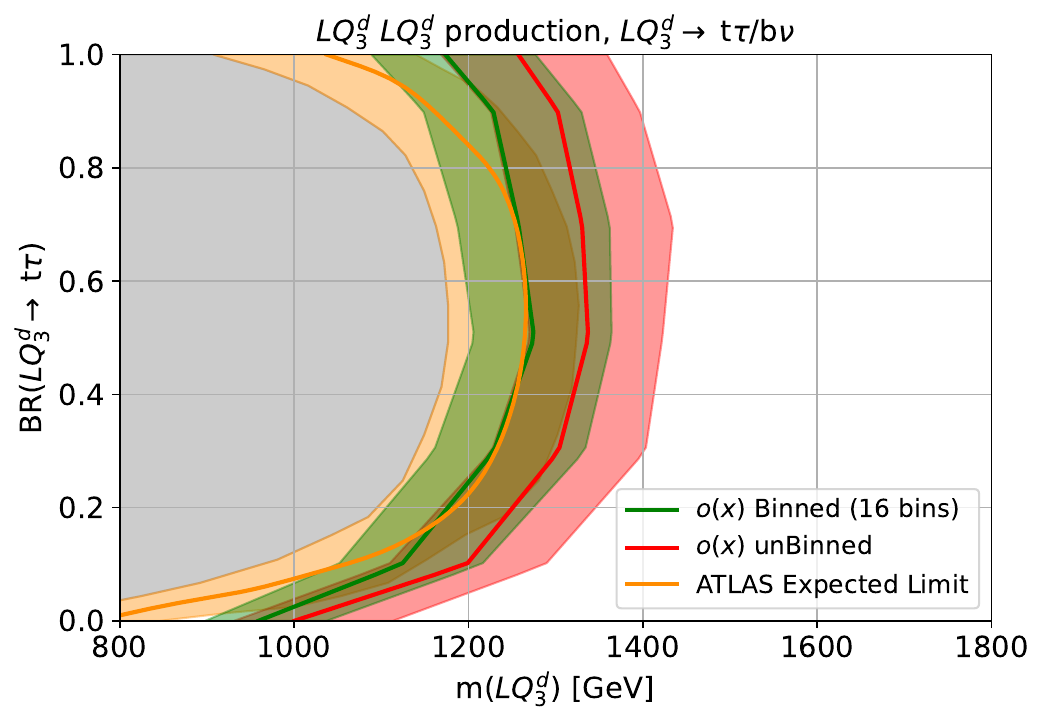}
\caption{Same as Fig.~\ref{LQ-nosys}, but including systematic uncertainties as detailed in the main text.}
\label{LQ-withsys}
\end{figure}

These results are shown in Fig.~\ref{LQ-withsys}. Comparing with the corresponding panels of Fig.~\ref{LQ-nosys}, we can see that the impact on the exclusion contours is only of a few per cent. Importantly, the effect in both methods is similar. This indicates that the treatment to include systematic uncertainties in the unbinned case provides a good numerical approximation despite not having an analytical expression as in the binned case, and it renders the limits shown in Fig.~\ref{LQ-nosys} stable. Finally, we have checked that including uncertainties in the other features does not impact significantly the results. However, we would like to remark that not all systematic uncertainties that originate from detector effects and theoretical assumptions were considered. Thus, although we expect a mild impact on the significance in the case these missing effects do not affect the most relevant variables in the ML discrimination, a full treatment including all sources and their correlations would be needed in a complete analysis.

\subsection{Prospects for 14-TeV LHC}
\label{projections}

\begin{figure}
  \centering
  \includegraphics[width=0.8\textwidth]{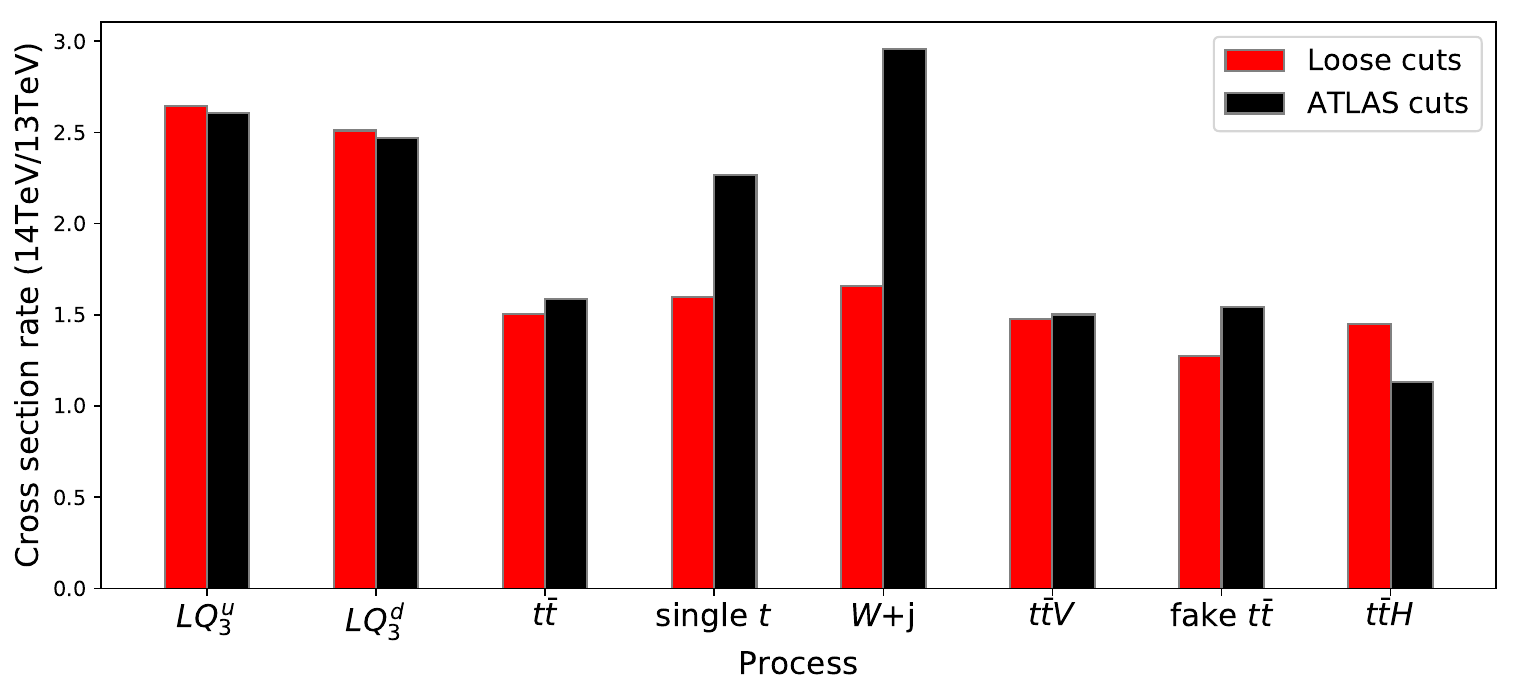}
\caption{Ratio of the cross-section at $\sqrt{s}=14$ TeV and $\sqrt{s}=13$ TeV for different processes: both types of LQs with $m(LQ_3^{u/d})=1200$ GeV and the main backgrounds considered in this work. Two sets of selection cuts are compared, in red the ones used throughout this work and in black the ones described by ATLAS~\cite{ATLAS:2021jyv}.}
\label{LQ-cross-14TeV}
\end{figure}

Next, we compare the expected exclusion limits obtained with $\sqrt{s}=13$ TeV and $\sqrt{s}=14$ TeV. We simulated new but small signal and background samples, $\sim$50k events per channel, with the same setup described in Section~\ref{simulation}, but at a center-of-mass energy of 14 TeV and with a HL-LHC ATLAS card for the {\sc Delphes} fast detector simulator. In Fig.~\ref{LQ-cross-14TeV} we show the ratio of the cross-section at $\sqrt{s}=14$ TeV and $\sqrt{s}=13$ TeV for different processes after selection cuts. For the `loose cuts', we can see that the cross-section of all the background increases by a similar factor, i.e. the hierarchy is the same for both center-of-mass energies. On the other hand, this is not true for the `ATLAS cuts', which involve cuts in high-level observables. Although the overall hierarchy in the background is conserved, the relative weight of the single-top and $W$+jets channels increases significantly.

Regarding the leptoquark signal, its cross section increases by a similar factor for both selection cuts. Importantly, the signal-to-background ratio is larger at 14 TeV which will impact significantly on the exclusion limit reach. We have checked this trend for all leptoquark masses and branching fractions.

\begin{figure}
  \centering
  \includegraphics[width=0.49\textwidth]{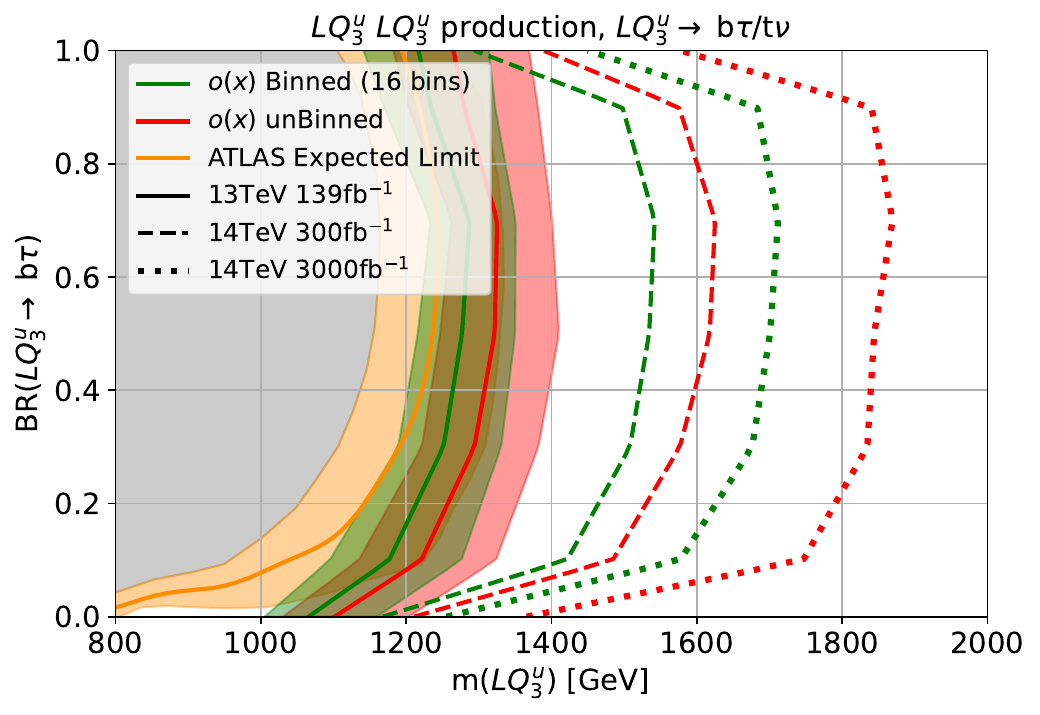}
  \includegraphics[width=0.49\textwidth]{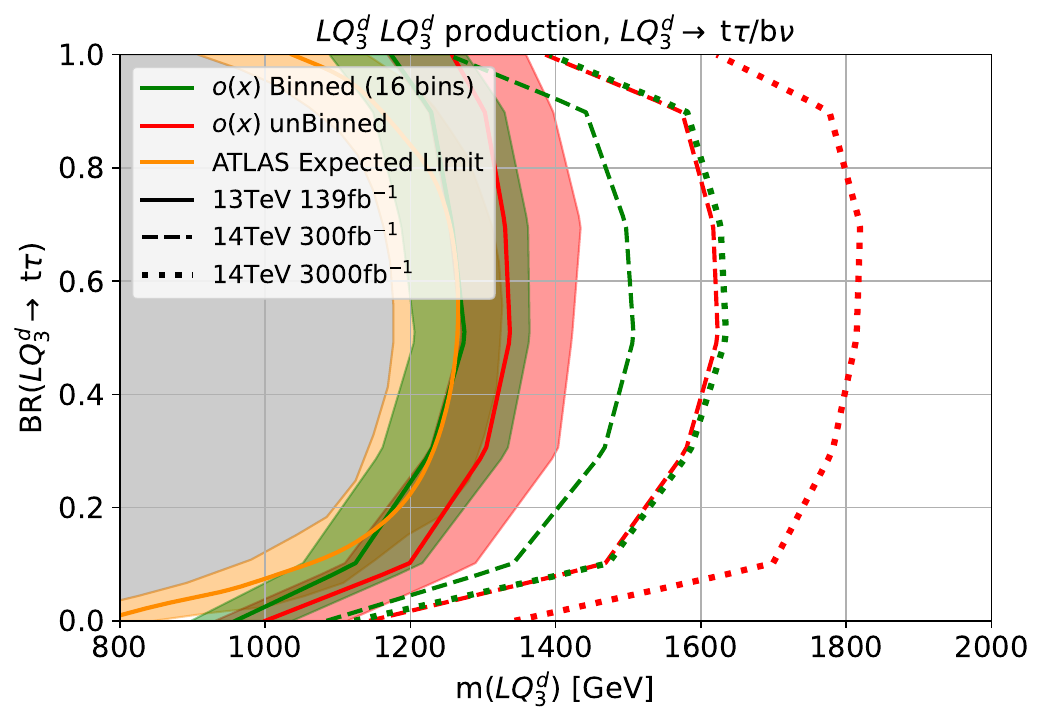}
\caption{Projected expected exclusion contours at the 95\% CL for $\sqrt{s}=13$ TeV and 139 fb$^{-1}$ (solid curves), $\sqrt{s}=14$ TeV and 300 fb$^{-1}$ (dashed curves), and $\sqrt{s}=14$ TeV and 3000 fb$^{-1}$ (dotted curves). All the results include systematic uncertainties as detailed in the main text. The rest of the references as in Fig.~\ref{LQ-nosys}.}
\label{LQ-withsys-14TeV}
\end{figure}

Since the generation of a new full set of events to train the ML algorithms is computationally very expensive, we employed the data sets at 13 TeV for the training stage, but used the cross-section values, relative weights of the background channels, relative weights of the LQs for different values of $\beta$, and expected number of signal and background events calculated at 14 TeV for the significance calculation in both binned and unbinned methods. Finally, in Fig.~\ref{LQ-withsys-14TeV} we present the projected expected exclusion contours at the 95\% CL for $\sqrt{s}=14$ TeV and 300 fb$^{-1}$ (dashed curves), and $\sqrt{s}=14$ TeV and 3000 fb$^{-1}$ (dotted curves). These results include systematic uncertainties, but for the sake of simplicity, we do not include the statistical uncertainty colored band. For comparison, we also include the limits for $\sqrt{s}=13$ TeV and 139 fb$^{-1}$ (solid curves) that were shown in Fig.~\ref{LQ-withsys}.

For $\sqrt{s}=14$ TeV and 300 fb$^{-1}$, the expected exclusion contours extend to masses $\sim$1.5 TeV and $\sim$1.6 TeV for the binned and unbinned methods, respectively, and for intermediate values of BR$(LQ_3^{u/d} \rightarrow q \ell)$. For $\sqrt{s}=14$ TeV and 3000 fb$^{-1}$ these are extended to $\sim$1.65 TeV and $\sim$1.8 TeV. As previously pointed out, the unbinned method provides the most stringent constraints. Remarkably, in the right panel, we can see that for $LQ^d_3$ the mass limit for the binned case at  $\sqrt{s}=14$ TeV and 3000 fb$^{-1}$ would be the same as the limit established by the unbinned method with 10 times less luminosity.

\section{Conclusions}
\label{conclu}

In this work, we have performed a collider analysis of LHC searches for pairs of scalar leptoquarks decaying into $b$-quarks and tau leptons. To carry out this phenomenological analysis, we have used an unbinned approach based on the so-called Machine-Learned Likelihoods method, in which we have incorporated as a novelty a simplified procedure for the inclusion of some systematic uncertainties. We remark that this method could be also applied to LQs coupled to other or mixed generations with promising results. However, it is not a goal of the present analysis to study these cases.

Our strategy employs a binary classifier that discriminates between signal and background, estimating their PDFs through the use of KDE. The fact of knowing the signal and background PDFs allows us to compute the likelihood function of the exclusion hypothesis test in order to impose 95\% CL exclusion limits on the parameter space defined by the LQ mass and its branching fraction into a third-generation quark and a third-generation charged lepton. The results with this unbinned method, for an LHC center-of-mass energy of 13 TeV and a luminosity of 139 fb$^{-1}$, seem to show a tendency towards a potential improvement of the exclusion limits set by the ATLAS analysis.

A first approach to the inclusion of systematic uncertainties is done by translating them from the physical-based space to the ML classifier output one. We have consider only individual 5-10\% uncertainties on the $\tau$-lepton $p_T$, $\MET$, and $H_T$, without correlations among them. The impact on the ML output was assessed then by replacing the training data set with variations of parameter values within their estimated uncertainties and repeating the analysis. Our results indicate that their impact on the exclusion limits is slight and the effect in both binned and unbinned methods is similar. Therefore, our approach to the inclusion of systematic uncertainties in the unbinned method provides a good numerical approximation despite the lack of an analytical expression as in the binned analysis. We have also checked that the inclusion of uncertainties in the other variables does not impact significantly on the results. Nevertheless, it is important to remark that a full treatment including all systematic sources and their correlations would be needed in a complete analysis. The simplified approach developed here attempts to show that the results obtained without the inclusion of any systematic uncertainty remain stable when at least a rough estimate of some of them is considered.

For the LHC at 13 TeV with 139 fb$^{-1}$ and intermediate branching fractions, we find exclusion limits for leptoquark masses $\sim$1.25 TeV and $\sim$1.3 TeV for the binned and unbinned methods, respectively. We have also estimated the prospects for the LHC at 14 TeV with luminosities of 300 fb$^{-1}$ and 3000 fb$^{-1}$. For the lower luminosity, the 95\% CL exclusion limits reach LQ mass values of $\sim$1.5 TeV and $\sim$1.6 TeV for the binned and unbinned methods, respectively. For 3000 fb$^{-1}$ these limits are extended to $\sim$1.65 TeV and $\sim$1.8 TeV, being the unbinned method the one that provides the most stringent constraints. 

\section*{Acknowledgments}
This work is partially supported by the ``Atracci\'on de Talento'' program (Modalidad 1) of the Comunidad de Madrid (Spain) under the grant number 2019-T1/TIC-14019 (EA, RMSS), and by the Spanish Research Agency (Agencia Estatal de Investigaci\'on) through the grants IFT Centro de Excelencia Severo Ochoa No CEX2020-001007-S (EA, RMSS, AP), PID2021-124704NB-I00 (EA, RMSS) and PID2021-125331NB-I00 (AP), funded by MCIN/AEI/10.13039/501100011033. AP also acknowledges support from the Comunidad Aut\'onoma de Madrid and Universidad Aut\'onoma de Madrid under grant SI2/PBG/2020-00005. DD and AS thank CONICET and ANPCyT (under project PICT 2018-03682). 

%%%%%%%
%%%%%%%
%%%%%%%

\bibliographystyle{JHEP}
\bibliography{biblio}
\end{document}